\documentclass[cit]{PoS}

\newcommand{\MeV}{\textrm{MeV}}
\newcommand{\GeV}{\textrm{GeV}}
\newcommand{\fm}{\textrm{fm}}
\newcommand{\fpi}{\ensuremath{f_{\pi}}}
\newcommand{\fK}{\ensuremath{f_{K}}}
\newcommand{\Dd}{\ensuremath{D^+}}
\newcommand{\Ds}{\ensuremath{D_s}}
\newcommand{\fDs}{\ensuremath{f_{\Ds}}}
\newcommand{\fDd}{\ensuremath{f_{\Dd}}}
\newcommand{\LamHQ}{\ensuremath{\Lambda_{\mathit{HQ}}}}
\newcommand{\comma}{,}
\newcommand{\fullStop}{.}

\title{The $\Ds$ and $\Dd$ Leptonic Decay Constants from Lattice QCD}

\ShortTitle{$\fDs$ and $\fDd$ from Lattice QCD}

\author{Fermilab Lattice and MILC Collaborations}
\author{%
A.~Bazavov$^a$,
C.~Bernard$^b$\thanks{cb@lump.wustl.edu},
C.~DeTar$^c$,
E.D.~Freeland$^{b}$,
E.~Gamiz$^{d,e}$,
Steven~Gottlieb$^{f,g}$,
U.M.~Heller$^h$,
J.E.~Hetrick$^i$,
A.X.~El-Khadra$^d$,
A.S.~Kronfeld$^e$,
J.~Laiho$^b$,
L.~Levkova$^c$,
P.B.~Mackenzie$^e$,
M.B.~Oktay$^c$,
M.~Di Pierro$^j$,
\speaker{J.N.~Simone}$^e$\thanks{simone@fnal.gov},
R.~Sugar$^k$,
D.~Toussaint$^a$,
and
R.S.~Van~de~Water$^l$ \\ \\
\llap{$^a$}Department of Physics, University of Arizona, Tucson, Arizona, USA \\
\llap{$^b$}Department of Physics, Washington University, St.~Louis, Missouri, USA \\
\llap{$^c$}Physics Department, University of Utah, Salt Lake City, Utah, USA \\
\llap{$^d$}Physics Department, University of Illinois, Urbana, Illinois, USA \\
\llap{$^e$}Fermi National Accelerator Laboratory, Batavia, Illinois, USA \\
\llap{$^f$}Department of Physics, Indiana University, Bloomington, Indiana, USA \\
\llap{$^g$}National Center for Supercomputing Applications, University of Illinois, Urbana, Illinois, USA \\
\llap{$^h$}American Physical Society, One Research Road, Box 9000, Ridge, New York, USA \\
\llap{$^i$}Physics Department, University of the Pacific, Stockton, California, USA \\
\llap{$^j$}School of Computer Sci., Telecom. and Info. Systems, DePaul University, Chicago, Illinois, USA \\
\llap{$^k$}Department of Physics, University of California, Santa Barbara, California, USA \\
\llap{$^l$}Physics Department, Brookhaven National Laboratory, Upton New York, USA \\
}

\abstract{%
We present the leptonic decay constants $\fDs$ and
$\fDd$ computed on the MILC collaboration's $2+1$ flavor asqtad
gauge ensembles. We use clover heavy quarks with the Fermilab
interpretation and improved staggered light quarks.  The simultaneous
chiral and continuum extrapolation, which determines both decay constants,
includes partially-quenched lattice results at
lattice spacings $a\approx0.09$, $0.12$ and $0.15$ fm.
We have made several recent improvements in our analysis:
a) we include terms in the fit describing leading order heavy-quark discretization effects,
b) we have adopted a more precise input $r_1$ value consistent with our other $D$ and $B$ meson studies,
c) we have retuned the input bare charm masses based upon the new $r_1$.
Our preliminary results are $\fDs=260\pm10\;\MeV$ and $\fDd=217\pm10\;\MeV$.

}

\FullConference{The XXVII International Symposium on Lattice Field Theory - LAT2009\\
		 July 26-31 2009\\
		 Peking University, Beijing, China}

\newcommand{\figCompareRone}{
\begin{figure}
\centering
\includegraphics[width=0.99\textwidth]{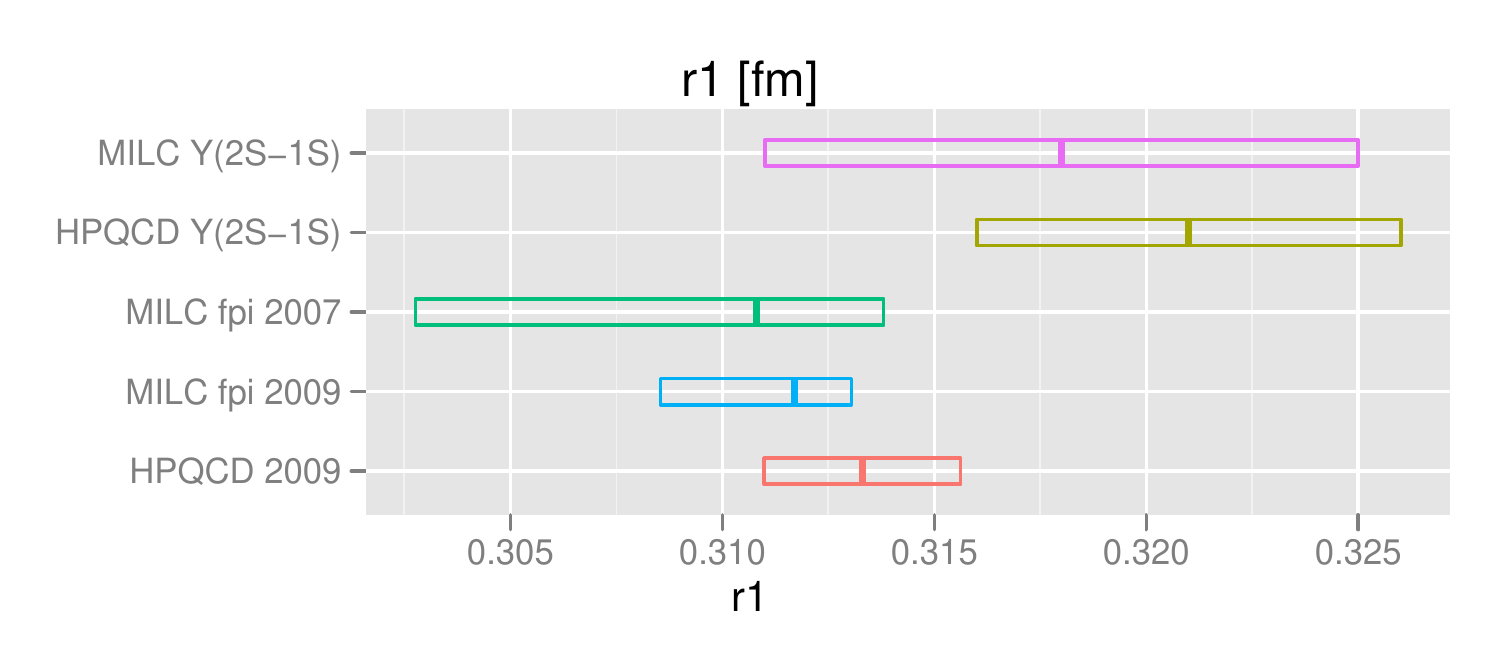}
\caption{%
Values of scale parameter $r_1$ in fermi units.
The ``HPQCD $\Upsilon(2S$-$1S)$'' value uses
the HPQCD collaboration $\Upsilon$ spectrum results to set the physical
value \cite{Gray:2005ur}. The ``MILC $\Upsilon(2S$-$1S)$'' value
derives from essentially the same spectrum analysis \cite{Aubin:2004wf}. MILC
determines $r_1$ more precisely from their calculation of $f_\pi$: ``MILC
$f_\pi$ 2007'' \cite{Bernard:2007ps} and ``MILC $f_\pi$ 2009'' \cite{Bazavov:2009tw}.
In a very recent update, ``HPQCD 2009'', several
physical quantities, including recent $\Upsilon$ results, are used
as inputs \cite{Davies:2009ts}.
}
\label{fig:rOneDeterminations}
\end{figure}}

\newcommand{\tabKappaTune}{
\begin{table}
\centering
\begin{tabular}{cc||ccc||ccc}\hline\hline
\multicolumn{2}{r||}{$r_1$ [fm]:}     &\multicolumn{3}{c||}{0.3108} &\multicolumn{3}{c}{0.318}  \\ \hline
  $a$     &$\kappa$ run  &$\kappa$ tune &$\delta\phi_s$ &\% $\phi_s$   &$\kappa$ tune &$\delta\phi_s$  &\% $\phi_s$ \\ \hline
 0.09   &0.127         &0.1272        &$-0.0043$       &$-0.56$         &0.1267        &$+0.0065$       &$+0.84$  \\
 0.12   &0.122         &0.1222        &$-0.0036$       &$-0.50$         &0.1215        &$+0.0091$       &$+1.26$  \\
 0.15   &0.122         &0.1222        &$-0.0031$       &$-0.42$         &0.1213        &$+0.0108$       &$+1.47$ \\ \hline
\multicolumn{2}{r||}{$\delta \fDs$ [MeV]} &\multicolumn{3}{c||}{$-1.8$} &\multicolumn{3}{c}{$+1.3$} \\ \hline
\end{tabular}
\caption{%
Tuning of $\kappa$ charm at the three lattice spacings for two choices
of $r_1$.  The shift $\delta\phi_s$ is the change in $\phi$ at the strange
quark mass when $\kappa$ changes from the run value to tuned $\kappa$ value.  The
corresponding change in extrapolated $\fDs$ is $\delta \fDs$.
In each case, all other extrapolation inputs are fixed to their
appropriate ($r_1$ dependent) values.}
\label{tab:kappaTuning}
\end{table}}

\newcommand{\figTheFIT}{
\begin{figure}
\centering
\includegraphics[width=0.92\textwidth]{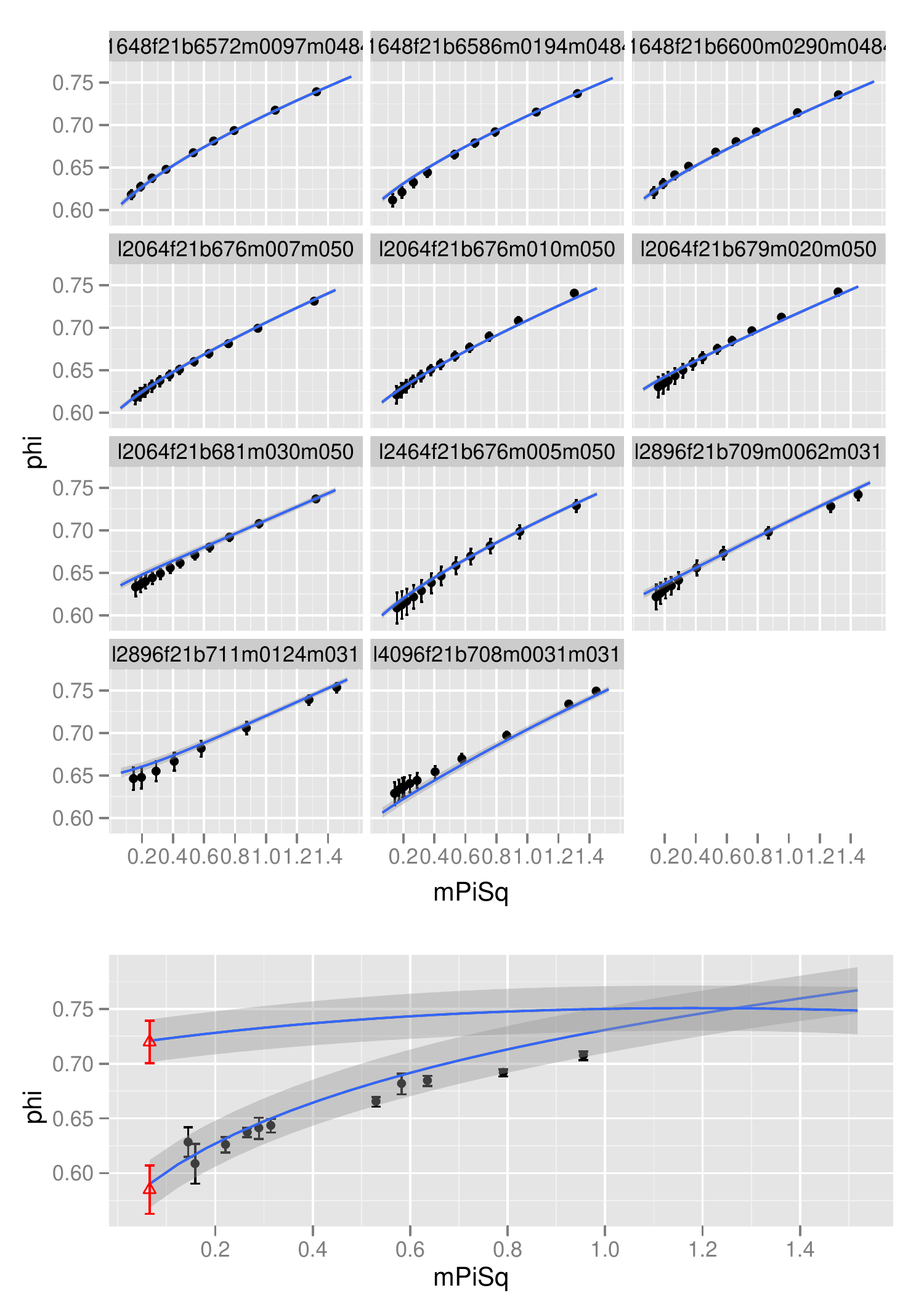}
\caption{%
The preliminary $D$ meson chiral extrapolation. The $3\times4$ matrix
of plots (top) show the $\phi$ data and corresponding fit including
$a^2$ effects. Reading from left to right and top to bottom, plots correspond to
$(a,m_l/m_h)$ values of $(0.15,0.2)$, $(0.15,0.4)$, $(0.15,0.6)$,
$(0.12,0.14)$, $(0.12,0.2)$, $(0.12,0.4)$, $(0.12,0.6)$, $(0.12,0.1)$,
$(0.09,0.2)$, $(0.09,0.4)$ and $(0.09,0.1)$.  The larger plot (bottom)
shows an overlay of the $\fDs$ and $\fDd$ extrapolations. The
extrapolated curves are the fit (with error bands) taking $a^2\to0$
and fixing/extrapolating the light quarks to their physical
masses. The extrapolations are not expected to go though any of the
points which are computed at finite $a$. None of the data points
having $m_q$ near $m_s$ seen the upper panel are visible in the $D_s$ extrapolation.}
\label{fig:fDfit}
\end{figure}}

\newcommand{\tabErrorBudget}{
\begin{table}
\centering
\begin{tabular}{ll@{\quad}l@{\quad}l} \hline\hline
                                 source &   $\phi_{\Ds}$ &   $\phi_{\Dd}$ &   $R_{{\Dd/\Ds}}$  \\ \hline
  statistics and discretization effects &            2.6 &            3.4 &           1.3  \\
                   chiral extrapolation &            2.0 &            2.5 &           0.8  \\
   inputs $r_1$, $m_s$, $m_d$ and $m_u$ &            0.7 &            0.7 &           0.3  \\
                            input $m_c$ &            1.0 &            1.2 &           0.2  \\
              $Z_V^{cc}$ and $Z_V^{qq}$ &            1.4 &            1.4 &             0  \\
              higher-order $\rho_{A_4}$ &            0.3 &            0.3 &           0.2  \\
                          finite volume &            0.2 &            0.6 &           0.6  \\
\hline
                                  total &            3.8 &            4.7 &           1.7  \\
\hline
\end{tabular}
\caption{%
Uncertainties as a percentage of $\phi$ and the ratio. The total combines all of the errors in quadrature.
}
\label{tbl:errorBudget}
\end{table}}

\newcommand{\figCompareDecayConstants}{
\begin{figure}
\centering
\includegraphics[width=0.98\textwidth]{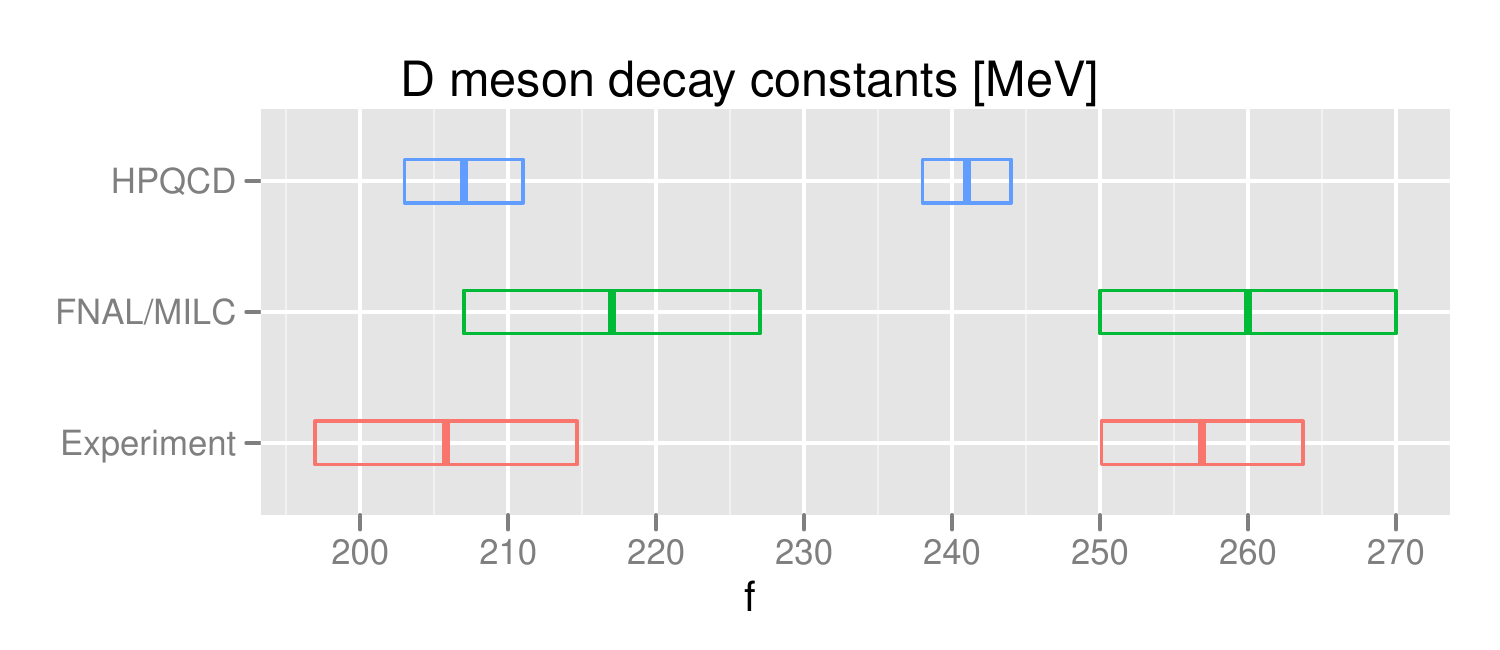}
\caption{%
Comparisons of the Fermilab/MILC values of $\fDd$ and $\fDs$ to
values from the HPQCD Collaboration \cite{Follana:2007uv} and recent
experimental values \cite{Eisenstein:2008sq}\cite{HFAG-Ds:2009xx}.
}
\label{fig:fDdeterimations}
\end{figure}}

\begin{document}

\section{Introduction}

We report on progress in the Fermilab Lattice and MILC
Collaboration calculation of the $D$ meson decay constants.  This work is a
continuation of the program that predicted the decay constants:
$\fDd=201(3)(17)$ and $\fDs=249(3)(16)\,\MeV$ \cite{Aubin:2005ar}, in
good agreement with the CLEO-c value of
$\fDd=205.8\pm8.5\pm2.5$ \cite{Artuso:2005ym,Eisenstein:2008sq}.  We
have since extended this calculation to two additional
ensembles at our finest lattice spacing $a\approx 0.09\;\fm$ and we have replaced a
limited set of very coarse ($a\approx 0.18\;\fm$) ensembles with higher
statistic ensembles at a somewhat finer spacing $a\approx 0.15\;\fm$.  In our
last update \cite{Bernard:2009wr} we reported: $\fDd=207(11)$ and
$\fDs=249(11)$, where $\fDs$ is about 0.6$\sigma$ lower than the
recent experimental average \cite{HFAG-Ds:2009xx}.  The value of $\fDs$
remains an pressing issue given that experimental average is about
2.1$\sigma$ higher than the most precise lattice result from the HPQCD
collaboration \cite{Follana:2007uv}. The apparent tension between
experiment and lattice predictions has motivated suggestions of
physics beyond the Standard Model \cite{Dobrescu:2008er}.

Smaller statistical uncertainties and better control of systematic
effects are key to resolving the $\fDs$ puzzle. In this
report, we have doubled statistics on the most chiral of the
$a\approx 0.09\;\fm$ lattices; otherwise, statistics have not changed. A new
generation of calculations, now underway, aims to increase
statistics by a factor of four overall. Our progress
includes: a) a better method of accounting for heavy-quark
discretization effects b) a more precise input value for
the scale parameter $r_1$, consistent with our other heavy quark studies
and c) more precisely tuned input charm kappa values.

\section{Staggered chiral perturbation theory for heavy-light mesons}
\label{sect:SXPT}

We use the asqtad improved staggered action for both sea and light valence
quarks.  Leading discretization effects split the light pseudoscalar
meson masses,
\begin{equation}
	M^2_{ab,\xi} = (m_a+m_b)\mu + a^2\Delta_\xi   \;\comma
        \label{eq.pionSplit}
\end{equation}
where there are sixteen tastes in representations $\xi=P, A, T, V, I$.

Staggered chiral perturbation theory for heavy-light mesons accounts
for such taste breaking effects \cite{Aubin:2005aq}.  At NLO in the
chiral expansion, for $2+1$ flavors, and at leading order in the heavy
quark expansion,
\begin{equation}
\phi_{H_q} = \Phi_0\left[1 + \Delta f_{H}(m_q,m_l,m_h) + p_H(m_q,m_l,m_h) + c_L\alpha_V^2a^2\right] \;\comma
\label{eq.SXPT-NLO}
\end{equation}
where $\phi_{H_q}=f_{H_q}\sqrt{m_{H_q}}$ and $f_{H_q}$ is the decay
constant of a heavy meson $H_q$ consisting of a heavy quark and a light quark of mass
$m_q$. The heavier sea quark has mass $m_h$ and the two degenerate light sea
quarks have mass $m_l$. The $\phi_{H_q}$, in general, are partially
quenched: $m_q \ne m_l$ and $m_q \ne m_h$.  The chiral logarithm terms,
$\Delta f_{H}$, are $a$ dependent as a consequence of the mass
splittings in Eq.~(\ref{eq.pionSplit}) as well as from ``hairpin''
terms proportional to the low energy constants $a^2\delta'_A$ and
$a^2\delta'_V$.  The $a$ dependence of the analytic terms, $p_H$,
ensures that $\phi_{H_q}$ is unchanged by a change in the chiral scale,
$\Lambda_\chi$, of the logarithms. The expression in
Eq.~(\ref{eq.SXPT-NLO}) is used in our combined chiral and continuum
extrapolations.  In practice, we add the NNLO analytic terms to the
fit function in order to extend the fit up to $m_q\sim m_s$ and
extract $\fDs$. Priors for the parameters $a^2\delta'_A$ and
$a^2\delta'_V$ as well as values of the physical light quark masses
are obtained from the MILC analysis of $\fpi$ and $\fK$ \cite{Bazavov:2009tw}.

\section{Discretization effects from clover heavy quarks}
\label{sect:HQdisc}

We use tadpole-improved clover charm quarks.  At leading order,
discretization errors are a combination of $\mathcal{O}(a^2\LamHQ^2)$ and
$\mathcal{O}(\alpha a \LamHQ)$ effects where $\alpha$ is the QCD coupling and
$\LamHQ$ is the scale in the heavy quark expansion. Our past studies
have estimated heavy quark discretization effects using such power
counting arguments to bound the error at the smallest lattice spacing,
taking $\LamHQ\approx 700\;\MeV$.  This rather crude method does not
effectively use the data to guide the error estimate.

This study introduces a new procedure: the leading order heavy quark
discretization errors are modeled to leading order as part of the
combined chiral and continuum extrapolation. At tree-level,
discretization effects arise from both the quark action and the
(improved) current \cite{Kronfeld:2000ck,Harada:2001fi,Oktay:2008ex}. We add five extra terms to
Equation~\ref{eq.SXPT-NLO}:
\begin{equation}
\Phi_0\left[a^2\LamHQ^2\left\{c_E f_E(am_Q)+c_Xf_X(am_Q)+c_Yf_Y(am_Q)\right\}+
\alpha_V a\LamHQ\left\{c_Bf_B(am_Q)+c_3f_3(am_Q)\right\}\right]
\label{eq.HQdiscTerms}
\end{equation}
The coefficients $c_E$, $c_X$, $c_Y$, $c_B$ and $c_3$ are additional
parameters determined in the fit while the $f_i$ are (smooth)
functions of the heavy quark mass, $am_Q$, known at tree level.  We
introduce priors for the coefficients constraining them to be $\mathcal{O}(1)$
while setting $\LamHQ=700\;\MeV$ and $m_c\sim1.2\;\GeV$.  Currently
the data are too noisy and the shapes of the functions $f_i$ are too
similar for the fit to prefer a particular $\LamHQ$.  Including the
heavy-quark discretization terms increases the decay constants by a
few MeV and increases the error from $\sim1.8\%$ to $\sim3.8\%$.  The
larger error now includes the residual heavy-quark discretization
uncertainty in addition to residual light-quark discretization effects
(encoded in Eq.~(\ref{eq.SXPT-NLO})) as well as statistical errors. We
quote a combined uncertainty from all the three sources of error.

\section{Lattice spacing determination from $r_1$}

\figCompareRone

The distance $r_1$ is a property of the QCD potential between heavy
quarks. The ratio $r_1/a$, for lattice spacing $a$, has been computed
for all of the MILC gauge ensembles. At intermediate stages of the
decay constant analysis quantities are converted from lattice units to
$r_1$ units using $r_1/a$. The value of $r_1$ must then be input in
order to convert results to physical units. The $r_1$ value is also an
input to the process of determining other quantities such as the
bare charm quark masses as discussed in the next section.

Figure~\ref{fig:rOneDeterminations} depicts several $r_1$
determinations.  The first two determinations historically (circa
2004--2005) are labeled ``HPQCD $\Upsilon(2S$-$1S)$''  \cite{Gray:2005ur} and ``MILC
$\Upsilon(2S$-$1S)$'' \cite{Aubin:2004wf}. They are both based on the same analysis of the
$\Upsilon$ spectrum by the HPQCD Collaboration using a subset of the
current MILC ensembles. The two determinations
differ mainly in the details of the continuum extrapolation.  The
MILC Collaboration is also able to infer a value of $r_1$ based on the
value of $\fpi$ they find in their analysis of the light mesons.
Recent light-meson analyses include results from finer lattice
spacings than the earlier $\Upsilon$ spectrum study and the resulting
$r_1$ values are known to better precision. The figure shows the result
of two recent analyses labeled `MILC $\fpi$ 2007''  \cite{Bernard:2007ps} and ``MILC $\fpi$
2009'' \cite{Bazavov:2009tw}.  The (preliminary) 2009 result agrees at the $0.9\sigma$ level
with the MILC $\Upsilon$ value but differs from the HPQCD $\Upsilon$
value at the $1.8\sigma$ level. As these proceedings were being
prepared, HPQCD published a new value for $r_1$ \cite{Davies:2009ts}, labeled ``HPQCD
2009'' in the figure, in much better agreement with MILC's recent $r_1$
values.

In this study, we use the MILC $r_1$ determinations from $f_\pi$ to set
the physical scale. Our central value for $r_1$ (the 2007 value) was
also used in our studies of the semileptonic decays on the same
lattices \cite{Bernard:2008dn,Bailey:2008wp}. The range of the 2009
MILC $r_1$ determination is used to set a symmetric uncertainty around
the central value. Hence, we take $r_1=0.3108\pm0.0022$.  Our previous
decay constant studies used the MILC $\Upsilon$ value:
$r_1=0.318\pm0.007$ as an input which is about one $\sigma$ higher.

\section{Retuning kappa charm}

\tabKappaTune

We determine the value of $\kappa$ for the charm quark by requiring
that the spin-averaged kinetic masses of the lattice pseudoscalar and
vector mesons made from a heavy clover quark and strange asqtad
valence quark equal the spin-averaged $D_s$ meson mass.  The tuning
depends upon $r_1$ in the conversion between lattice and physical
masses.

In the past year, we have conducted new kappa-tuning runs with at least
four times the statistics of our older tunings. At each lattice
spacing, we simulated mesons for three values of $\kappa$ around charm
and three light-quark masses around strange allowing us to retune
$\kappa$ for a given $r_1$.

Table~\ref{tab:kappaTuning} shows preliminary tunings for $\kappa$
charm based upon the two input values: $r_1=0.3108\;\fm$ (present
value) and $r_1=0.318\;\fm$ (past studies). For each $r_1$, the
(preliminary) tuned kappa and the corresponding change
$\delta\phi_s=\phi_s(\kappa\;\mathit{tune})-\phi_s(\kappa\;\mathit{run})$
is listed by lattice spacing. The ``run'' kappa values are those used
in the decay constant simulations.  We adjust each $\phi_q$ point by
$\delta\phi_s$ prior to the chiral extrapolation to correct for the
mistuning of kappa.  The bottom row of the table shows the resulting
change in $\fDs$.  The opposite signs of the differences show that
keeping kappa tuned partly compensates the change in $r_1$.  We find
that changing $r_1$ from $0.318\;\fm$ to $0.3108\;\fm$ while keeping
kappa charm tuned increases $\fDs$ by about $4.2\;\MeV$.

\section{The chiral and continuum extrapolation, results and uncertainty budget}

\figTheFIT

We fit $\phi_{D_q}$ results from lattice simulations on eleven asqtad
MILC ensembles  \cite{Aubin:2004wf} at the three lattice spacings:
$a\approx 0.09$, $0.12$ and $0.15\;\fm$. Our valence quark masses are in the
range $0.1m_s \le m_q \lesssim m_s$.  Since our last report, we have
doubled the statistics at the most chiral of the $a\approx 0.09$
ensembles. The $3\times4$ panel of plots at the top in
Fig.~\ref{fig:fDfit} shows the $\phi_{D_q}$ points and the fit where
the fit function includes the lattice-spacing effects described in
Sections~\ref{sect:SXPT} and \ref{sect:HQdisc}.  The plot at the bottom of
Fig.~\ref{fig:fDfit} shows the extrapolations in the limit $a=0$.  The
upper ($\Ds$) curve shows $m_l\to\hat{m}$ setting $m_q=m_h=m_s$, while
the lower ($\Dd$) curve shows $m_q,m_l\to\hat{m}$ setting $m_h=m_s$.
The physical quark mass inputs are from the MILC light meson analysis
and $\hat{m}=(m_u+m_d)/2$.  The points denoted by the red triangles
correspond to physical $\fDs$ and $\fDd$.  Our preliminary results
are:
%
% central value Fit 8
%
\begin{equation}
\begin{array}{c@{,\quad}c@{\quad\textrm{and}\quad}c}
\fDd = 217 \pm 10 \; \MeV
&\fDs = 260 \pm 10 \; \MeV
&\fDs/\fDd = 1.20 \pm 0.02 \;\fullStop
\end{array}
\end{equation}

\tabErrorBudget

We have combined the statistical and the systematic uncertainties
listed in Table~\ref{tbl:errorBudget} in quadrature. Our largest
uncertainty is the combined uncertainty from statistical and residual
discretization effects. The second largest uncertainty, chiral
extrapolation, is an estimate of chiral expansion effects not included
in the fit function and effects from variation in the extrapolation
procedure. The third largest error is the statistical error in the
nonperturbative calculation of the current renormalizations $Z_V^{cc}$
and $Z_V^{qq}$.  The value of $\fDs$ is about eleven MeV (one sigma)
higher than our earlier value. Using nominal kappa values rather than
tuned values at the previous $r_1$ value accounts for about 1.3 MeV of
the difference.  Changing to the new $r_1$ while keeping kappa tuned
results in a 4.2 MeV increase.  Incorporating heavy quark effects into
the fit increases $\fDs$ by about 2 MeV.  Higher statistics on the most
chiral of the $a\approx 0.09\;\fm$ lattice increases $\fDs$ by about 1
MeV.  These changes combine nonlinearly in the fit to yield the net
increase.

%
% TODO: kappa_charm error: double differences between Fit 5 and Fit 1: 1.0% (Ds), 1.2% (Dd) and 0.2% (ratio)
%
%

\figCompareDecayConstants

Figure~\ref{fig:fDdeterimations} compares the Fermilab and MILC
Collaboration values for the decay constants with the HPQCD
Collaboration \cite{Follana:2007uv} values and with the experimental
results. The experimental result for $\fDd$ is from
CLEO \cite{Eisenstein:2008sq} while the $\fDs$ value is the Heavy
Flavor Averaging Group average \cite{HFAG-Ds:2009xx} of determinations
by CLEO, BaBar and Belle. The Fermilab / MILC results remain in agreement
with experiment. The total error on the experimental average for $\fDs$
is now smaller that our error providing a challenge for future lattice
determinations. The apparent discrepancy between the HPQCD value of
$\fDs$ and the other two $\fDs$ values is most striking.  The HPQCD
value is lower by about 1.8--2.1$\sigma$. The source of this
difference may be clarified by further lattice simulations.

\section{Summary and future plans}

We have made several improvements in our analysis: a)
discretization effects from both heavy and light quarks are modeled
in our extrapolation function, b) we adopted a more precise
$r_1$ value which derived from the MILC
$\fpi$ analysis rather than the $r_1$ value related to early $\Upsilon$
spectrum results c) we have improved the tuning of kappa charm. These
improvements to the analysis will be more crucial in our next
generation of decay constant study. We will
increase statistics by a factor of four and extend the analysis to the
finer lattice spacings $a\approx 0.06$ and $0.045\;\fm$ which
will reduce our combined statistical
plus discretization error as well as help reduce uncertainties attributed
to chiral extrapolation procedures. In addition, a new high-statistics
computation of the nonperturbative part of the current renormalization
aims for an error below the 0.5\% level.

%\figTheFIT

%\bibliographystyle{JHEP}%{unsrt}
%\bibliography{bib/FermilabAndMILC,bib/HPQCD,bib/MILC,bib/Experiment}

\begin{thebibliography}{10}

\bibitem{Aubin:2005ar}
C. Aubin {\em et al.}, {\it {Charmed meson decay constants in three-flavor
  lattice QCD}},  {\em Phys. Rev. Lett.} {\bf 95} (2005) 122002,
  [\href{http://xxx.lanl.gov/abs/hep-lat/0506030}{{\tt hep-lat/0506030}}].

\bibitem{Artuso:2005ym}
{\bf CLEO} Collaboration, M. Artuso {\em et al.}, {\it {Improved Measurement of
  $\mathcal{B}(D^+ \to \mu^+ \nu)$ and the Pseudoscalar Decay Constant
  $f_{D^+}$}},  {\em Phys. Rev. Lett.} {\bf 95} (2005) 251801,
  [\href{http://xxx.lanl.gov/abs/hep-ex/0508057}{{\tt hep-ex/0508057}}].

\bibitem{Eisenstein:2008sq}
{\bf CLEO} Collaboration, B. I. Eisenstein {\em et al.}, {\it {Precision
  Measurement of $\mathcal{B}(D^+ \to \mu^+ \nu)$ and the Pseudoscalar Decay
  Constant $f_{D^+}$}},  {\em Phys. Rev.} {\bf D78} (2008) 052003,
  [\href{http://xxx.lanl.gov/abs/0806.2112}{{\tt 0806.2112}}].

\bibitem{Bernard:2009wr}
C. Bernard {\em et al.}, {\it {B and D Meson Decay Constants}},  {\em PoS} {\bf
  LATTICE2008} (2008) 278, [\href{http://xxx.lanl.gov/abs/0904.1895}{{\tt
  0904.1895}}].

\bibitem{HFAG-Ds:2009xx}
H. F. A. G. C. Physics), ``$f_{D_s}$ world average.''
  \href{http://www.slac.stanford.edu/xorg/hfag/charm/PIC09/f_ds/results.html}{%
www.slac.stanford.edu/xorg/hfag/charm/PIC09/f\_ds/results.html}, 2009.

\bibitem{Follana:2007uv}
{\bf HPQCD} Collaboration, E. Follana, C. T. H. Davies, G. P. Lepage, and
  J. Shigemitsu, {\it {High Precision determination of the $\pi$, $K$, $D$ and
  $D_s$ decay constants from lattice QCD}},  {\em Phys. Rev. Lett.} {\bf 100}
  (2008) 062002, [\href{http://xxx.lanl.gov/abs/0706.1726}{{\tt 0706.1726}}].

\bibitem{Dobrescu:2008er}
B. A. Dobrescu and A. S. Kronfeld, {\it {Accumulating evidence for nonstandard
  leptonic decays of $D_s$ mesons}},  {\em Phys. Rev. Lett.} {\bf 100} (2008)
  241802, [\href{http://xxx.lanl.gov/abs/0803.0512}{{\tt 0803.0512}}].

\bibitem{Aubin:2005aq}
C. Aubin and C. Bernard, {\it {Staggered chiral perturbation theory for
  heavy-light mesons}},  {\em Phys. Rev.} {\bf D73} (2006) 014515,
  [\href{http://xxx.lanl.gov/abs/hep-lat/0510088}{{\tt hep-lat/0510088}}].

\bibitem{Bazavov:2009tw}
{\bf The MILC} Collaboration, A. Bazavov {\em et al.}, {\it {Results from the
  MILC collaboration's $SU(3)$ chiral perturbation theory analysis}},  {\em
  PoS} {\bf LAT2009} (2009) 079, [\href{http://xxx.lanl.gov/abs/0910.3618}{{\tt
  0910.3618}}].

\bibitem{Kronfeld:2000ck}
A. S. Kronfeld, {\it {Application of heavy-quark effective theory to lattice
  QCD. I: Power corrections}},  {\em Phys. Rev.} {\bf D62} (2000) 014505,
  [\href{http://xxx.lanl.gov/abs/hep-lat/0002008}{{\tt hep-lat/0002008}}].

\bibitem{Harada:2001fi}
J. Harada {\em et al.}, {\it {Application of heavy-quark effective theory to
  lattice QCD. II: Radiative corrections to heavy-light currents}},  {\em Phys.
  Rev.} {\bf D65} (2002) 094513,
  [\href{http://xxx.lanl.gov/abs/hep-lat/0112044}{{\tt hep-lat/0112044}}].

\bibitem{Oktay:2008ex}
M. B. Oktay and A. S. Kronfeld, {\it {New lattice action for heavy quarks}},
  {\em Phys. Rev.} {\bf D78} (2008) 014504,
  [\href{http://xxx.lanl.gov/abs/0803.0523}{{\tt 0803.0523}}].

\bibitem{Gray:2005ur}
A. Gray {\em et al.}, {\it {The Upsilon spectrum and $m_b$ from full lattice
  QCD}},  {\em Phys. Rev.} {\bf D72} (2005) 094507,
  [\href{http://xxx.lanl.gov/abs/hep-lat/0507013}{{\tt hep-lat/0507013}}].

\bibitem{Aubin:2004wf}
C. Aubin {\em et al.}, {\it {Light hadrons with improved staggered quarks:
  Approaching the continuum limit}},  {\em Phys. Rev.} {\bf D70} (2004) 094505,
  [\href{http://xxx.lanl.gov/abs/hep-lat/0402030}{{\tt hep-lat/0402030}}].

\bibitem{Bernard:2007ps}
C. Bernard {\em et al.}, {\it {Status of the MILC light pseudoscalar meson
  project}},  {\em PoS} {\bf LAT2007} (2007) 090,
  [\href{http://xxx.lanl.gov/abs/0710.1118}{{\tt 0710.1118}}].

\bibitem{Davies:2009ts}
C. T. H. Davies, E. Follana, I. D. Kendall, G. P. Lepage, and C. McNeile, {\it
  {Precise determination of the lattice spacing in full lattice QCD}},
  \href{http://xxx.lanl.gov/abs/0910.1229}{{\tt 0910.1229}}.

\bibitem{Bernard:2008dn}
C. Bernard {\em et al.}, {\it {The $\bar{B} \to D^{*} \ell \bar{\nu}$ form
  factor at zero recoil from three-flavor lattice QCD: A Model independent
  determination of $|V_{cb}|$}},  {\em Phys. Rev.} {\bf D79} (2009) 014506,
  [\href{http://xxx.lanl.gov/abs/0808.2519}{{\tt 0808.2519}}].

\bibitem{Bailey:2008wp}
J. A. Bailey {\em et al.}, {\it {The $B \to \pi \ell \nu$ semileptonic form
  factor from three-flavor lattice QCD: A Model-independent determination of
  $|V_{ub}|$}},  {\em Phys. Rev.} {\bf D79} (2009) 054507,
  [\href{http://xxx.lanl.gov/abs/0811.3640}{{\tt 0811.3640}}].

\end{thebibliography}

\end{document}